\documentclass[useAMS,usenatbib]{mn2e}
\usepackage{txfonts,graphicx}

\title[MAGIC upper limits on the GRB~090102 afterglow]{MAGIC upper limits on the GRB~090102 afterglow}

\author[J. Aleksi\'c et al.]
{
\parbox{\textwidth}{
\small{
J. Aleksi\'c$^{1}$, 
S. Ansoldi$^{2}$, 
L. A. Antonelli$^{3}$\thanks{Corresponding author}, 
P. Antoranz$^{4}$, 
A. Babic$^{5}$, 
P.~Bangale$^{6}$, 
U. Barres de Almeida$^{6}$, 
J. A. Barrio$^{7}$, 
J. Becerra Gonz\'alez$^{8}$, 
W. Bednarek$^{9}$
% ...
}}\vspace{0.4cm}\\
Affiliations can be found at the end of the article.}

\begin{document}

\date{Accepted 1988 December 15. Received 1988 December 14; in original form 1988 October 11}

\pagerange{\pageref{firstpage}--\pageref{lastpage}} \pubyear{2002}

\maketitle

\label{firstpage}

\begin{abstract}
Indications of a GeV component in the emission from GRBs are known since the EGRET observations during the 1990's and they have been confirmed by the data of the \textit{Fermi} satellite. These results have, however, shown that our understanding of GRB physics is still unsatisfactory. The new generation of Cherenkov observatories and in particular the MAGIC telescope, allow for the first time the possibility to extend the measurement of GRBs from several tens up to hundreds of GeV energy range. Both leptonic and hadronic processes have been suggested to explain the possible GeV/TeV counterpart of GRBs. Observations with ground-based telescopes of very high energy photons (E$>$30 GeV) from these sources are going to play a key role in discriminating among the different proposed emission mechanisms, which are barely distinguishable at lower energies. MAGIC telescope observations of the GRB~090102 ($z=1.547$) field and \textit{Fermi} Large Area Telescope (LAT) data in the same time interval are analyzed to derive upper limits of the GeV/TeV emission. We compare these results to the expected emissions evaluated for different processes in the framework of a relativistic blastwave model for the afterglow. Simultaneous upper limits with \textit{Fermi} and a Cherenkov telescope have been derived for this GRB observation.  The results we obtained  are compatible with the expected emission although the difficulties in predicting the HE and VHE emission for the afterglow of this event makes it difficult to draw firmer conclusions.  Nonetheless, MAGIC sensitivity in the energy range of overlap with space-based instruments (above about 40 GeV) is about one order of magnitude better with respect to \textit{Fermi}. This makes evident the constraining power of ground-based observations and shows that the MAGIC telescope has reached the required performance to make possible GRB multiwavelength studies in the very high energy range.
\end{abstract}


\section{Introduction}
\label{sec:magicobs}

Since the discovery of Gamma-Ray Bursts (GRB)  in the late 1960's \citep{klebesadel73}, these energetic and mysterious  phenomena have been targets of large observational efforts. The discovery of their afterglow in late '90 \citep{Costa97,vanparadijs97} provided a great boost in GRB studies at all wavelengths. The wealth of available information put severe constraints on the various families of interpretative scenarios, showing an unexpected richness and complexity of possible behaviours \cite[see e.g.][]{Geh09}. The first observations at MeV-GeV energies with the Energetic Gamma-Ray Experiment Telescope (EGRET) on board the Compton Gamma-Ray Observatory \cite[CGRO; ][]{Hurley94,Dingus95}, showed that the high energy (HE: 1 MeV-30 GeV ) and very high energy range (VHE: 30 GeV - 30 TeV) can be powerful diagnostic tools for the emission processes and physical conditions of GRBs. The launch of \textit{Fermi} \citep{Band09}, with its Large Area Telescope \cite[LAT;][]{Atwood09}, showed that, at least for the brightest events, GeV emission from GRBs is a relatively common phenomenon \citep{Granot10}. However, a satisfactory interpretative framework of the GeV emission is still lacking. In this context, ground-based imaging atmospheric Cherenkov telescopes (IACTs), such as MAGIC\footnote{http://wwwmagic.mppmu.mpg.de/index.en.html}, H.E.S.S.\footnote{http://www.mpi-hd.mpg.de/hfm/HESS} and VERITAS\footnote{http://veritas.sao.arizona.edu}, despite the reduced duty cycle of ground-based facilities, provide access to the $\sim$100 GeV to TeV energy interval for GRB observations. Furthermore, the energy range down to $\sim80$ GeV, which was accessible almost exclusively with space-based instruments, has been opened to ground-based observations by the MAGIC observatory \citep{Aliu08,Sch10}. Together with the multiwavelength coverage provided by the LAT instrument, this makes possible the complete coverage of the 1-100 GeV energy range with the advantage, in the VHE domain, of an increase of $\sim$ 2-3 order of magnitude in the sensitivity relative to space-based instruments. Moreover, the low energy trigger threshold of MAGIC makes less relevant the effect of the source distance.  The flux above $\sim$100 GeV is attenuated by pair production with the lower energetic (optical/IR) photons of the diffuse Extragalactic Background Light \cite[EBL; ][]{Nikishov62,GS66}. The resulting cosmic opacity to VHE $\gamma$-rays heavily affects Cherenkov observations, especially for GRBs which are sources with an average redshift slightly larger than 2 \citep{Fynbo09}. Therefore, the higher the redshift, the lower the likelihood of detection at a given energy (i.e., about a $90\%$ of flux reduction at 100 GeV for a $z=2$ source following \citet[][]{Alberto11}). In addition, the transient and unpredictable nature of GRBs makes it difficult for large ground-based instruments such as IACTs to point them rapidly enough to catch the prompt emission and the early afterglow phases, when these sources are expected to be observable at high energies \citep[see e.g.][for a discussion about IACTs perspectives for GRB observations]{Covino09b}. MAGIC has the advantage, compared to the other IACTs, in its low-energy sensitivity and pointing speed \cite[e.g.][]{Garcz09}. Several attempts to observe GRB emission have been discussed \citep{MAGICGRB1,MAGICGRB2,MAGICCovino10}. In all cases only upper limits have been derived. Similar results have also been reported by other IACTs \citep{Tam06,Ah09,veritas_grb}. As discussed above, the two most limiting factors are the high-redshift of the source and the delay of the observation.\\

\begin{figure}
 \includegraphics[width=88mm]{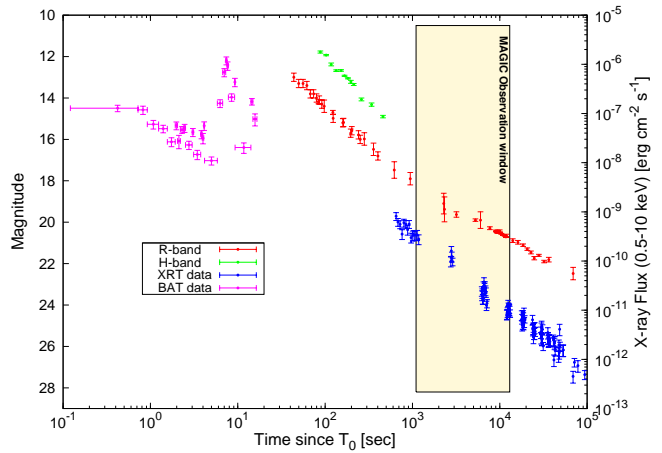}
 \caption{Light curve for GRB~090102. Data in the R-band (red points) were taken from the TAROT, REM, NOT, GROND and Palomar telescopes. The early near-infrared H band (blue points) are from the observations of the REM telescope. All magnitudes are expressed in the Vega system. X-ray data are the unbinned {\it Swift}/XRT and BAT data in the 0.5-10 keV (green and magenta point respectively). The MAGIC observation window is also plotted. R and H data from \citet[][]{Gendre10}. XRT and BAT data retrieved from http://www.swift.ac.uk/burst\_analyser   \citep{Evans10}.}
 \label{fig:mwl}
\end{figure}

In this paper we report and discuss the MAGIC observation of GRB~090102, a GRB at a redshift about $1.5$ observed at low zenith angle and good weather conditions. These observations permitted data-taking with an energy threshold of about 30 GeV. However, no gamma-ray signal was detected and hence only upper limits could be derived.\\
Section 2 gives general information about GRB~090102. In Sections 3 and 4 we discuss the MAGIC and LAT data sample respectively, while sections 5 and 6 introduce and develop the interpretative scenario. In Sect. 7 we evaluate the effect of the EBL absorption on the lowest energy bins allowed by our observation and finally, we discuss our results in a general theoretical scenario in the last section.\\
We assume a $\Lambda{\rm CDM}$ cosmology with $\Omega_{\rm m} = 0.27$, $\Omega_\Lambda = 0.73$ and $h_0 = 0.71$. At the redshift of the GRB the proper distance is $\sim 4.5$\,Gpc ($\sim 1.38 \times 10^{28}$\,cm). Throughout this paper the convention $Q_x = Q/10^x$ has been adopted in CGS units.

\section[]{GRB~090102}

GRB~090102 was detected and located by the \textit{Swift} satellite \citep{Geh04} on January 2$^{\rm {nd}}$, 2009 at 02:55:45 UT \citep{Mangano09} and also by the Fermi-{\it GBM} detector. The prompt light curve was structured in four partially overlapping peaks \citep{Saka09} for a total T$_{90}$ of $27.0 \pm 2.0$s. Since the burst was also detected by \textit{Konus Wind} \citep{Golenetskii09} and \textit{Integral} \citep{Mangano09b}, it has been possible to obtain a very good reconstruction of the prompt emission spectral parameters. The time-averaged spectrum can be modeled with the classical Band function \citep{Band93} with peak energy E$_{peak}$=451$^{ \ +73}_{ \ -58}$ keV and a total fluence in the 20 keV - 2 MeV range of 3.09$^{ \ +0.29}_{ \ -0.25} \times $ 10$^5$ erg cm$^{-2}$ \citep{Golenetskii09}. Early optical follow-up measurements were performed by many groups like TAROT \citep{Klotz09} at T$_0$+40.8s, the REM robotic telescope at T$_0$+53s \citep{Covino09} and GROND telescope \citep{Afonso09} at T$_0$+2.5h. Optical spectroscopy was rapidly obtained with the NOT telescope by \cite{deug09}. They found evidence of several absorption metal lines, including Fe~II, Mg~II, Mg~I, Al~II, Al~III and C~IV, at a common redshift of $z=1.547$. The resulting isotropic energy value  $E_{iso}=5.75 \times 10^{53}$ erg and the rest frame peak energy $E_{peak}$=1149$^{ \ +186}_{ \ -148}$ keV are in good agreement with the Amati relation \citep{Amati09}.
The multiwavelength light curve is shown in Fig.~\ref{fig:mwl} in which, data in the R and H band correspond to a rest-frame UV and optical emission respectively. X-ray data are the unbinned \textit{Swift}/XRT and BAT data in the 0.5-10 keV. According to \citet{Gendre10}, it is very difficult to model the whole afterglow in a standard scenario (see next section). Moreover, it showed a distinct behaviour in the optical and in X-rays. The X-ray light curve showed an uninterrupted decay from about $400$s from the GRB onset up to $5\times10^6$s, when {\it Swift} ceased observations of the event. The optical light-curve, monitored from several tens of seconds to slightly more than a day from the T$_0$, showed a steep-to-shallow behavior with a break at about 1ks. Before the break, the optical flux decay index is $\alpha_1=1.50 \pm 0.06$ while the index becomes $\alpha_2=0.97 \pm 0.03$ after the break, steeper and flatter respectively, when compared to the simultaneous X-ray emission. This behavior strongly resembles that showed by GRB~061126 \citep[][]{Gomboc08,Perley08} and GRB~060908 \citep{Covino10}.\\
Other observations were performed at later time by GROND \citep[T$_0$+2.5h -][]{Afonso09}, Palomar \citep[T$_0$+50 min.-][]{palomar} and IAC80 \citep[T$_0$+19.2h -][]{iac80} telescopes while during the following days, the NOT \citep{Malesani09} and HST \citep{Levan09} provided the detection of the host galaxy. In the radio energy band, the VLA \citep{Chandra09} and the Westerbork Synthesis Radio Telescope (WSRT) \citep{Horst09} performed follow-up observation at 8.46 GHz and 4.9 GHz with no afterglow detection and upper limits evaluation. A detailed discussion of the follow up observations for this burst can be found in \citet{Gendre10}.

\section{MAGIC follow up observation and analysis}

The MAGIC telescope located at Roque de los Muchachos (28.75$^{\circ}$N, 17.89$^{\circ}$W, La Palma, Canary Islands) performed a follow-up measurement of GRB~090102. The data presented in this paper were taken when MAGIC was operating as a single telescope. The MAGIC telescope was autonomously repointed and started the observations at T$_0$ +255 s, following the GRB alert from Fermi-{\it GBM}. Later on, the shift crew operating the telescope realized that the GBM coordinates (RA: $08^{{\rm h}}$~$35^{{\rm m}}$~$06^{{\rm s}}$; DEC: $37^{\circ}$~ $16^{\prime}$~$48^{\prime \prime}$) differed from the BAT coordinates (RA: $08^{{\rm h}}$~$33^{{\rm m}}$~$02^{{\rm s}}$; DEC: $33^{\circ}$~ $05^{\prime}$~$29^{\prime \prime}$) by more than 4 degrees. Consequently, the telescope was repointed to the BAT coordinates and re-started observations by T$_0$+1161 s. After this burst, the alert system was modified to cope with this situation. First data runs were taken at very low zenith angles from 5$^{\circ}$ reaching 52$^{\circ}$ at the end of data taking at 06:54:01 UT after 13149 s of observation. MAGIC upper limits above 80 GeV have already been published for this GRB \citep{Gaug09}, while results and scientific discussion about a subsequent dedicated analysis focused in the low energy band \citep{Gaug09b} will be presented here. To ensure the lowest energy threshold, only data taken with zenith distance $<$ 25$^{\circ}$, corresponding to the first 5919 s of observation (data sub-sample up to 04:53:32 UT) have been taken into account during this analysis. By employing the MAGIC-1 sum trigger system \citep{Aliu08}, an analysis threshold of around 30 GeV is achieved, which is evaluated from MC simulations. In order to accurately estimate the background from hadronic atmospheric showers, an OFF data sample was taken one night later with the telescope pointing close to the burst location and in the same observational conditions and instrument setup. Data were analyzed using the MAGIC Analysis and Reconstruction Software \citep[MARS; ][]{MARS,MARS2} and processed using the standard Hillas parameters \citep{Hillas}. $\gamma$/hadron separation and energy estimation were performed using a multi-dimensional classification method \citep[Random Forest; ][]{RF} while arrival directions  of the $\gamma$ photons is reconstructed using the DISP algorithm \citep{DISP}. The {\it alpha} parameter is then used to evaluate the significance of the signal in six energy bins. In spite of the low energy analysis threshold, no significant excess of gamma-ray photons have been detected from a position consistent with GRB~090102. Differential upper limits (ULs) assuming a power-law $\gamma$-ray spectrum with spectral index of $\Gamma=-2.5$ and using the method of \citet[][]{rol05} were evaluated with a 95$\%$ confidence level CL  and 30$\%$ estimation of systematic uncertainties and are reported in Tab.~\ref{tab:ul} and Fig.~\ref{fig:ul}. 

\begin{table}
\begin{center}
\caption[UL Table]{MAGIC-I 95$\%$ confidence level upper limits for the afterglow emission of GRB~090102. The values correspond to the first 5919 s of observation from 03:14:52 to 04:53:32 UT.$\ \ ^{\alpha}$ Bins central energy was evaluated applying all analysis cuts to MC simulations. $\ \ ^{\beta}$  Statistical significance of the excess events observed by MAGIC.}
\begin{tabular}{|c|c|c|c|}\hline
{\bfseries E bin} &  {\bfseries $<$E$>^{\ \alpha}$} & {\bfseries $\sigma^{\ \beta}$} & {\bfseries Average Flux Limits} \\
$[{\rm GeV}]$ & $[{\rm GeV}]$&  & $[{\rm erg \ cm}^{-2} \ {\rm s}^{-1}]$\\
\hline
25 - 50 & 43.9 & 0.83 & $8.7 \times 10^{-10}$ \\
50 - 80 & 57.3 & -0.30 & $1.5 \times 10^{-10}$ \\
80 - 125 & 90.2 & 1.09 & $3.1 \times 10^{-10}$ \\
125 - 175 & 137.2 & 0.51 & $2.2 \times 10^{-10}$ \\
175 - 300 & 209.4 & 0.90 & $1.6 \times 10^{-10}$ \\
300 - 1000 & 437.6 & -0.48 & $0.3 \times 10^{-10}$ \\
\hline
\end{tabular}
\end{center}
\label{tab:ul} 
\end{table}

\begin{table}
\begin{center}
\caption[Breaks Table]{Spectrum break energies for the different considered processes. The value of the expected SSC emission in the first MAGIC energy bin ($\sim$40 GeV) is also shown with and without considering the EBL absorption.  We refer to \cite{Zhang01} for the numerical results presented in this paper.} 

\begin{tabular}{|ccc|}\hline
{\bfseries Synchrotron (e)}& {\bfseries{SSC}} &{\bfseries{Synchrotron (p)}} \\
\hline
E$_{m} \simeq 0.6$ eV & E$_m^{\rm ssc} \simeq 1.1$ MeV & E$_m^p \simeq 10^{-8} $eV \\
E$_{c} \simeq 4.1$ eV & E$_c^{\rm ssc} \simeq 47$ MeV & E$_c^p \simeq 140 $ TeV \\
E$_{max} \simeq 207$ MeV & E$_{\rm KN} \simeq 60 $ TeV & E$_{max}^p \simeq $ 1.7 MeV \\
$\approx 5 \times 10^{-11}$& $ \approx 1.1 \times 10^{-10}$ & $ \approx 4\times 10^{-17}$ \\
-  & $4.3 \times 10^{-11} \ (3.4 \times 10^{-11})$  &  - \\
\hline
\end{tabular}
\end{center}
\label{tab:Breaks} 
\end{table}

\section{LAT observation and analysis}
\label{sec:lat}

\begin{figure}
\centering
 \includegraphics[width=88mm]{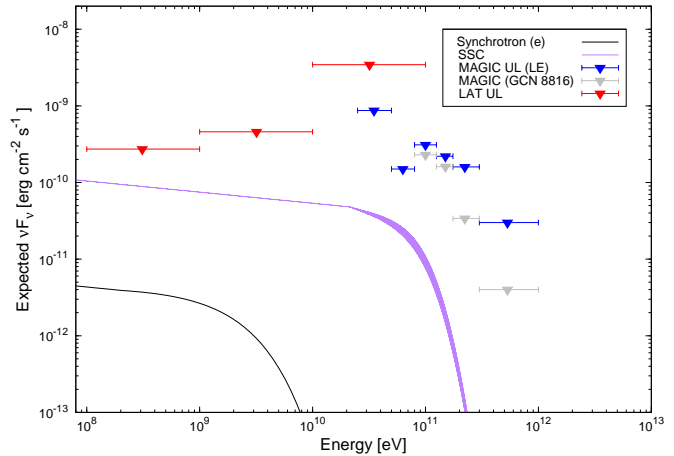}
 \caption{SSC modeled emission during the afterglow of GRB~090102. Blue triangles are 95\% CL upper limits derived by MAGIC for low energy (LE) analysis. The relatively more constraining upper limit in the 50-80 GeV is due to a negative significance energy bin. For comparison, the regular energy range MAGIC ULs \citep[][]{Gaug09} are also reported in light grey.  The red triangles report the \textit{Fermi}-LAT 95\% CL upper limits. The purple and black curves depict the expected energy flux according to the GRB afterglow model described in sec.~\ref{sec:he} and sec.~\ref{sec:lowe}.  Physical parameters are $\epsilon_e=0.1$, $\epsilon_B=0.01$, $E_{52}=4.5$ and $T=T_0+4$ks at a redshift $z=1.547$. The shaded region shows the uncertainty in the EBL absorption, as prescribed in~\citet[][]{Alberto11}.}
\label{fig:ul}
\end{figure}

The {\it Fermi} observatory is operating in a sky survey mode and the {\it Swift} localization of GRB~090102 was observable by the LAT instrument approximately $3300$ sec after trigger and remained within the LAT field-of-view ($\Theta_{boresight} \lesssim 60^{\circ}$) for a duration of $\sim 2300$ sec. We analyzed the Fermi-LAT data using the Science Tools 09-30-01 with $Pass7V6$ 'Source' event class. We used the publicly-available models for the Galactic and isotropic diffuse emissions, $gal\_2yearp7v6\_trim\_v0.fits$ and $iso\_p7v6source.txt$, that can be retrieved from the Fermi Science Support Center~\footnote{http://fermi.gsfc.nasa.gov/ssc/data/access/lat/BackgroundModels.html}. No significant excess was found in this observation, so we computed upper limits in 3 different energy bands: [0.1-1 GeV], [1- 10 GeV], and [10-100 GeV].  We first fit the broad energy range (from 0.1 to 100 GeV) using the unbinned likelihood analysis, which was then used to constrain the background model. Then we froze the normalizations of the isotropic and Galactic diffuse templates, and independently fit the source in the 3 different energy bands, using the unbinned profile likelihood method to derive 95\% LAT upper limits. The following upper-limit values were derived for the [0.1 - 1 GeV], [1 - 10 GeV], [10 - 100 GeV] energy ranges respectively: $2.73 \times 10^{-10}$, $4.58 \times 10^{-10}$, $3.45 \times 10^{-9}$ ${\rm erg \ \ cm}^{-2} {\rm  \ \ s}^{-1}$ and are depicted in Fig.~\ref{fig:ul}. These ULs are more constraining than the ones reported in \citet[][]{Inoue13}. The reason for that is the usage of $P7V6$ 'Source' instead of $P6V3$ 'Diffuse', and also the usage of a different procedure to parametrize the diffuse background in the three differential energy bins. Even if observed with a considerable time delay, the achieved energy threshold of MAGIC permits a better overlap with LAT in the GeV range when compared with previous results on GRB by MAGIC and other IACTs. Thus, it has been possible to derive simultaneous upper limits with a complete coverage of the energy range from 0.1 GeV up to TeV using MAGIC and {\it Fermi}-LAT. Furthermore, it is worth stressing that, in the energy range where the two instruments overlap (range [25-100 GeV]), the upper limits derived by MAGIC are about one order of magnitude lower than those from {\it Fermi}-LAT. 

\section{The low energy scenario}
\label{sec:lowe}

In a commonly accepted scenario \citep[see e.g][for a review]{Meszaros06}, GRB dynamics during the prompt phase are governed by relativistic collisions between shells of plasma emitted by a central engine (internal shocks). Similarly, the emission during the afterglow is thought to be connected to the shocks between these ejecta with the external medium (external shocks). Several non thermal mechanisms, indeed, have been suggested as possible sources of HE and VHE\footnote{GRBs show their phenomenology mainly in the X-ray and soft $\gamma$-ray energy band (1 keV - 1 MeV). To avoid confusion with the {\it Fermi}-LAT and IACT operational energy range ($>$20 MeV and $>$25 GeV, respectively ), we will refer to the former as a "low energy" range.} photons. They include both leptonic and hadronic processes, \citep[see e.g. for a review][]{Zhang01,Gupta07,FanSUMMARY08,Ghisellini10}. In the most plausible scenario, electron synchrotron radiation is the dominant process in the low energy regime. Within this scenario, the GRBs spectra are usually approximated by a broken power law in which the relevant break energies are the minimum injection $\nu_{m}$ and the cooling $\nu_{c}$. The first one refers to emission frequency of the bulk of the electron population (where most of the synchrotron emission occurs), while the cooling frequency identifies where electrons effectively cool. Both are strongly dependent on the microphysical parameters used to describe the GRB environment and, for a constant density $n$ of the circumburst diffuse interstellar medium (ISM) they are given by \citep{Zhang01}: 

\begin{eqnarray}
\nu_m = 8.6\times10^{17} \left( \frac{p-2}{p-1} \right)^2 \left(\frac{\epsilon_e}{\zeta_e} \right)^2 t_h^{-3/2} E_{52}^{1/2} \epsilon_B^{1/2} (1+z)^{1/2} \ \ \ \ [{\rm Hz}]
\label{eq:num}
\end{eqnarray}
\begin{eqnarray}
\nu_{\rm c} = 3.1\times10^{13} \left( 1+{Y_e} \right)^{-2} \epsilon_B^{-3/2} E_{52}^{-1/2} n^{-1} t_h^{-1/2} (1+z)^{1/2} \ \ \  [{\rm Hz}]
\label{eq:nuc}
\end{eqnarray}

\indent where $\epsilon_e$ and $\epsilon_B$ are the energy equipartition parameter for electrons and magnetic field, $E_{52}$ is the energy per unit solid angle, $t_h$ is the observer's time in hours, $\zeta_e$ is fraction of the electrons that enter in the acceleration loop and $Y_{\rm e}$ is the ratio between synchrotron and Inverse Compton (IC) cooling time, known as Compton factor \citep[see e.g.][]{SaEs01,PaKu00}. As a matter of fact, we have explicitly assumed that the contribution of the Compton scattering is not negligible in the afterglow at the considered time and, as a consequence, the cooling break is reduced by a factor (1+$Y_e$). It is important to remark that the change in slope of the optical decay observed in GRB~090102 suggests that the standard model cannot adequately describe the dynamics of this event. The steep-to-shallow behaviour could be interpreted as due to a termination shock, locating the end of the free-wind bubble generated by a massive progenitor at the position of the optical break. However, it is also possible to hypothesize that the early steeper decay is simply due to the superposition of the regular afterglow and a reverse shock present only at early times. It is not our purpose to analyze and discuss the several physical scenarios that are proposed to describe the afterglow, so we continue to model the burst emission assuming the afterglow could be described in the standard context of a relativistic shock model.

\section{Modeling the VHE emission}
\label{sec:he}

Any attempt to a meaningful modeling of the possible VHE emission component both during the prompt emission and the afterglow,  must rely on information coming from the low energies \citep[see e.g.][]{MAGICCovino10}. At the same time, the modeling of the low-energy afterglow can furthermore help in limiting the intrinsic degeneracy or even, to some extent, arbitrariness in the choice of the various possible HE and VHE afterglow parameters.  Following \citet{Gendre10} we assume that the cooling frequency at the time of MAGIC observation is located between optical and X-ray band. Thus, we can estimate the slope of the energy particles distributions which is correlated with the optical decay index. With the observed optical spectral index of $0.97 \pm 0.03$ \citep{Gendre10}, we obtain a value for $p$ from the relation $\frac{4}{3}(p-1)=0.97$ of $p=2.29 \pm 0.04$ in good agreement with numerical simulations which suggest a value of $p$ ranging between 2.2-2.3 \citep{Acht01,Vie03}. We will assume that at the time of the MAGIC observation, the outflow expands into a diffuse medium with a constant density of the order $n\sim$1  cm$^{-3}$ and we will further assume that all electrons are accelerated in the shocks ($\zeta_{e} \sim 1$). At the same time, from the available data, we can only constrain the values of $\epsilon_{B}$ and $\epsilon_{e}$. Assuming that the optical light curve time break \citep{Melandri10} is less than the start time of the shallow decay phase (T$_{\rm break} \lesssim  10^3$s), we obtain $ 0.04 \lesssim \epsilon_{e} \lesssim 0.2$ and $ 7 \times 10^{-4} \lesssim \epsilon_{B} \lesssim 0.05$ which only barely fix the $\epsilon_{B}$,$\epsilon_e$ values. We thus assume, within these limits, $\epsilon_{B} \sim 0.01$ and $\epsilon_{e} \sim 0.1$ which correspond to typical values for the late afterglow \citep{PaKu02,Yos03}. The most plausible process producing VHE photons is the IC mechanism in the variant of Synchrotron Self Compton (SSC) \citep{Zhang01,SaEs01}. Within this process, the low energy photons produced in the standard synchrotron emission are the seed photons that are pushed into the VHE band by IC scattering. Similarly to the previous case, the predicted SSC spectrum is characterized by the two typical frequencies $\nu_{m}^{ssc} = \gamma_{m}^2 \nu_{m}$ and $\nu_{c}^{ssc}=\gamma_{c}^2 \nu_{c}$ where $\gamma_{m}$ and $\gamma_{c}$ are the Lorentz factor for the electrons of frequencies $\nu_{m}$ and $\nu_{c}$. Since electrons are ultrarelativistic ($\gamma_{{m,c}} \sim 10^3$), SSC radiation can easily reach the GeV-TeV domain. Following \cite{Zhang01} and according to Eqs.\,(\ref{eq:num}) and (\ref{eq:nuc}) we have: 
\begin{eqnarray}
\nu_m^{ssc} = 1.3&\times&10^{22} \ \ \left(\frac{\epsilon_e}{\zeta_e} \right)^4 \left( \frac{p-2}{p-1} \right)^4 \nonumber \\
                                &\times& E_{52}^{3/4} n^{-1/4} t_h^{-9/4} \epsilon_B^{1/2} (1+z)^{5/4} \ \ \ \ \ \ \ \ \ \ \ \ \ [{\rm Hz}]
\end{eqnarray}
\begin{eqnarray}
\nu_c^{ssc} = 1.2&\times&10^{25} (1+Y_e)^{-4} \nonumber \\
                                &\times& E_{52}^{-5/4} n^{-9/4} \epsilon_B^{-7/2} t_h^{-1/4} (1+z)^{-3/4} \ \ \ \ \ \ \ \ [{\rm Hz}]
\end{eqnarray}

\noindent while the expected maximum flux density is \citep{Zhang01}:

\begin{eqnarray} 
F_{\nu, max}^{ssc}= 17 \ \ \ \zeta_e^2 (n E_{52})^{5/4} t_h^{1/4} (1+z)^{3/4} \ \ \ [{\rm nJy}]
\end{eqnarray}

Basically, the new spectral feature has the same shape of the underlying synchrotron component with a new break in the spectrum ($E_{KN}$) due to the decreasing of the IC cross-section with energy \citep{Fragile04}.  However,  this cut-off is found to be above few tens of TeV in our case, securing that MAGIC upper limits stay below this limit.  

We also consider proton synchrotron emission as an hadronic-originated component \citep{BoDe98,PeWa05}.  Although protons are poor emitters with respect to the electrons due to their high mass, they can be accelerated in internal or external shocks in the same way as electrons producing synchrotron radiation. However, the peak flux for the two particle emission components is determined by the mass ratio $\approx \frac{m_e}{m_p}$ which implies that hadronic component is usually much smaller with respect to the electron emission. Nevertheless, while electrons cool quickly, protons cooling times are much longer since

\begin{eqnarray}
\frac{\nu_{\rm c,p}}{\nu_{c,e}}=\left( \frac{1+Y_{e}}{1+Y_{p}} \right)^2 \left( \frac{m_{p}}{m_{e}} \right)^6
\label{eq:coolp}
\end{eqnarray} 

where ${\rm Y}_{p}=\frac{\sigma_{{p}\gamma}}{\sigma_{p,T}} {\rm Y}_{e}$ and $\sigma_{{p}\gamma}$, $\sigma_{p,T}$ are the proton-gamma interaction and Thomson cross section respectively. We note that since $\sigma_{{p}\gamma} >> \sigma_{p,T}$, $Y_p >> 1$ and proton's energy is mostly lost in p-$\gamma$ interaction rather than synchrotron emission. However, Eq.\ref{eq:coolp} implies that in the same cases proton synchrotron emission can exceed the electron component in the MAGIC energy band. In Sec.\ref{sec:concl}, we will briefly discuss the relative importance of the above described emission components.

\begin{figure}
 \centering
 \includegraphics[width=88mm]{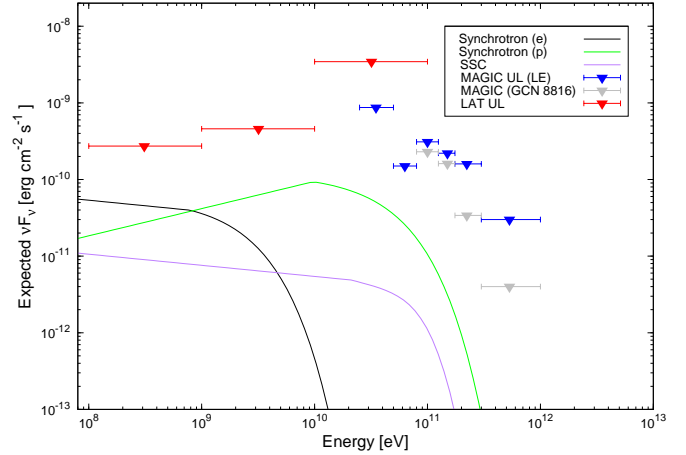}
 \caption{Modeled SED in Hadronic-dominated scenario for GRB afterglow. Used parameters are $E_{52}=10^3$, $T_{obs}=T_{0}+4$~ks, $\epsilon_{e}=10^{-3}$, $\epsilon_{B}=0.01$, $n=100$ cm$^{-3}$.}
 \label{fig:ul2}
\end{figure}

\section{Extragalactic background light attenuation}
\label{sec:ebl}
Gamma ray absorption by pair production with EBL plays a key role in VHE astronomy since it significantly limits the IACTs capability in detecting sources at redshift z$>$1. The optical depth $\tau$ is strictly connected to the light content of the Universe and the source distance. In the past years, several EBL models have been proposed providing a wide range of values for $\tau$ from 1 up to 6 for a z$\sim$1 source at 100 GeV \citep[see e.g.][]{Kneiske04,Stecker06,Stecker08}, which gives an attenuation in the expected flux ranging between $1/3$ to more than $1/100$. However, the more recent EBL models \citep[][]{Franceschini08,Alberto11}, although based on different assumptions, are converging to stable results. Within this context, \citet{Alberto11} have used real data on the evolution of the galaxy population taken from the AEGIS catalogue to evaluate EBL intensity for a wide range of redshift. The reliability of the results have been tested on the three most distant object observed by MAGIC \citep{Alberto11b}. Moreover, the EBL intensity evaluated using this model matches the minimum level allowed by galaxy counts which leads to the highest transparency of the universe to VHE $\gamma$-rays. We used the model of \citet{Alberto11} to evaluate the EBL absorption obtaining a value for $\tau$ of $0.218^{\ +0.075}_{\ -0.041}$ at about 40 GeV. This gives an attenuation of the flux at the same energy of the order of $\sim20\%$, a value that does not significantly compromise detection capability of MAGIC. However, the optical depth increases quickly with energy reaching the values of $\sim 1.5$ and $\sim 14.4$ at 100 GeV and 500 GeV respectively and this makes necessary to lower the energy threshold of the observation. In the case of GRB\,090102, MAGIC shows its capability to perform observation at very low energy limiting the $\gamma$ absorption even for moderate redshift sources.

\section{Discussion}
\label{sec:concl}

Although only upper limits have been obtained, the possibility of having simultaneous observations with {\it Fermi}-LAT in the energy range 0.1-100 GeV and MAGIC in the energy range that starts at 25 GeV (hence overlapping with LAT) make the GRB\,090102 a good case study in spite of its relatively high redshift. However, it has to be remarked that  GRB\,090102 can be considered as a common GRB in terms of both energetics and redshifts. Higher expected fluxes can be foreseen in case of more energetic events that are not so rare accordingly to {\it Fermi} results. We have used the equations of relativistic shock model in order to predict, in a reliable way, the expected VHE emission in the LAT and MAGIC energy range. From numerical results it is evident that for the chosen parameters and at our observation time, leptonic components are the dominant mechanisms from the low to the very high energy. Following Eq.(\ref{eq:coolp}), the cooling frequency for protons is usually located well above the very high energy range ($>$TeV) and this make the process potentially interesting for MAGIC observations. However, to make the two emissions comparable in the low energy regime, and the proton component to dominate the leptonic one at high energies, a fine tuning in the parameters choice is needed implying  $\frac{\epsilon_e}{\epsilon_p} \approx \frac{m_e}{m_p} \approx 10^{-3}$. Similar results can be obtained for the IC component. In both cases, however, a higher total energy release of $\approx10^{55}$ erg and a circumburst density medium of $\approx100$ cm$^{-3}$ are needed to maintain the low energy flux at the observed level. This makes the possibility of observing the hadronic emission component with the MAGIC telescope unrealistic, at least for a canonical model. A sketch of the scenario described above is shown in Fig.\ref{fig:ul2}.\\

Here, we did not take into account other hadronic-induced processes such as $\pi^0$ decay \citep{BoDe98}. However, it has been shown that they could have a non negligible effect at higher energies. For our parameters, the SSC process looks the most reliable mechanism in the VHE range. Indeed, we obtain $E_{\rm m}^{ssc} \sim 1$ MeV and $E_{\rm c}^{ssc} \sim 50$ MeV. We conclude that MAGIC observation ($> 40$ GeV) was carried out in the SED region where $\nu$F$_{\nu} \propto \nu^{(2-{\rm p})/2}$ \citep[][]{WeiFan07} so that it is possible to evaluate the expected SSC emission. Following \citet[][]{Zhang01}:

\begin{eqnarray}
\nu F_{\nu} = \nu F_{\nu,max} {\nu_{c,ssc}}^{1/2} {\nu_{m,ssc}}^{(p-1)/2} \nu^{(2-p)/2}
\end{eqnarray}

\begin{figure}
 \centering
 \includegraphics[width=88mm]{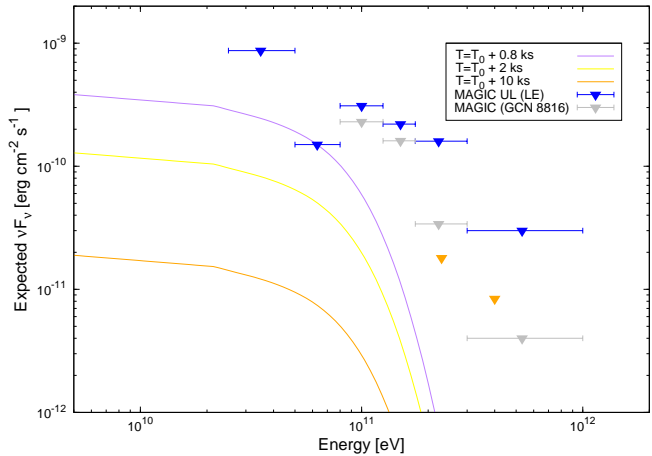}
 \caption{The expected SSC emission at different time. $T_0+0.8$ ks (purple curve), $T_0+2$ ks (yellow curve), $T_0+10$ ks (orange curve). In the latter case the corresponding VERITAS $99\%$ upper limits on $\sim 5000$ s of observation have been plotted \citep{veritas_grb} for comparison.}
 \label{fig:sens}
\end{figure}

which gives (for MAGIC first energy bin and taking into account the EBL absorption) $\nu F_{\nu, ssc}$(40 GeV)$ \approx 3.4 \times 10^{-11}$ erg cm$^{-2}$ s$^{-1}$. This result lies about one order of magnitude below the corresponding upper limits. However, a change in the microphysical parameters can influence the VHE emission giving scenarios with substantially higher flux. One of the most critical variables is the intensity of the magnetic field, which influences the relative importance between synchrotron and SSC emission. It is of particular importance for this event, since the most striking observational feature of GRB\,090102 was the observation of $\sim$10$\%$ polarization in the optical at early times \citep[about 3 min after the GRB; ][]{Steele09}. One of the most plausible interpretations is that the outflow generating the GRB is driven by a large scale ordered magnetic field, which generates polarized optical synchrotron emission in the optical observable during the reverse shock phase \citep{Steele09}. Large values of the magnetic field affect in a significant way the HE emission since it reduces the importance of the IC component. However, the regular forward-shock emission should not be affected by this ordered magnetic field \citep{Covino07,Mundell07}. The time-scale of the MAGIC observations are indeed likely late enough not to require this further parameter in the modeling. Such a delay, in association with the moderate source distance, militate against performing constraining observations with MAGIC. Indeed, we estimated that $\nu F_{\nu} \propto t^{-1.2}$. This implies that lowering the temporal delay of the observations can make the expected emission higher by one order of magnitude as illustrated in  Fig.\ref{fig:sens} where the expected SSC emission at different time is showed. 

\section{Future prospects}
\label{sec:future}

\begin{figure}
\centering
 \includegraphics[width=88mm]{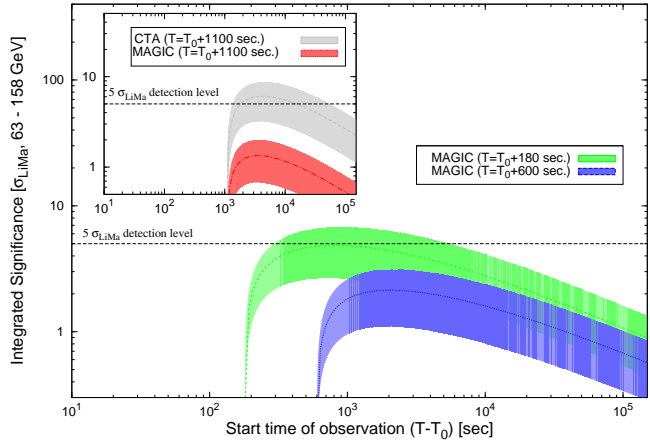}
\caption{Integrated $\sigma_{LiMa}$ significance as a function of observation time in the 63-158 GeV for the MAGIC stereo system in the case of a GRB event similar to the one reported in this paper. Significance curve evolutions are showed for different starting of observation times after GRB onset: 180 sec. (green), 600 sec. (blue), 1100 sec. (red). In this latter case (inner plot), the foreseen performance for CTA assuming the preliminary sensitivity achieved with MC simulations in the same energy range is showed.The colored areas shows the assumed $50\%$ systematic errors in the effective area evaluation as explained in~\citep[][]{Sav13}.}
\label{fig:prospect}
\end{figure}

Catching VHE signal from GRBs is one of the primary target of the MAGIC telescope and future IACTs like the Cherenkov Telescope Array (CTA).  Our estimates show that for this particular GRB, MAGIC follow-up observations made within the first 1-2 minutes from the trigger time would have the potential to detect the VHE component or at least to evaluate constraining upper limits (see Fig.\ref{fig:prospect}). This demonstrates both the capabilities of the system and the necessity of a fast-response observations. 

\noindent As GeV emission is found to be relatively common in \textit{Fermi} GRBs \citep[see e.g][]{LATGRB1}, the unique opportunity of having simultaneous follow-up with LAT and the MAGIC telescope will make accessible the end of the electromagnetic spectrum of GRBs and will have an important role in constraining different emission mechanisms and the space parameters. Moreover, the recent technical improvement of the Stereo MAGIC system \citep{magic_performance}, will bring an improvement in the instrument sensitivity in its low energy range. The steeper decay of the flux makes in any case difficult late time ($>200$s) detections for such moderate high-redshift event (see Fig.\ref{fig:prospect}). On the other hand, such a timescale is  well within the pointing capabilities of the present generation of IACTs (e.g. MAGIC) that are able to perform follow-up measurements within few hundreds of seconds. Basing on the preliminary sensitivity of the future CTA\footnote{obtained from http://www.cta-observatory.org/ctawpcwiki \\/index.phpWP\_MC\#Interface\_to\_WP\_PHYS}, a detection will instead be possible, within the assumed model, even on later time ($>1000$s) and higher redshift events. 

\section*{Acknowledgments}

We acknowledge an anonymous referee for useful comments\\

We would like to thank the Instituto de Astrof\'{\i}sica de Canarias for the excellent working conditions at the Observatorio del Roque de los Muchachos in La Palma. The support of the German BMBF and MPG, the Italian INFN, the Swiss National Fund SNF, and the Spanish MICINN is gratefully acknowledged. This work was also supported by the CPAN CSD2007-00042 and MultiDark CSD2009-00064 projects of the Spanish Consolider-Ingenio 2010 programme, by grant DO02-353 of the Bulgarian NSF, by grant 127740 of the Academy of Finland, by the DFG Cluster of Excellence ``Origin and Structure of the Universe'', by the DFG Collaborative Research Centers SFB823/C4 and SFB876/C3, and by the Polish MNiSzW grant 745/N-HESS-MAGIC/2010/0.\\

The {\it Fermi}-LAT Collaboration acknowledges support from a number of agencies and institutes for both development and the operation of the LAT as well as scientific data analysis. These include NASA and DOE in the United States, CEA/Irfu and IN2P3/CNRS in France, ASI and INFN in Italy, MEXT, KEK, and JAXA in Japan, and the K.~A.~Wallenberg Foundation, the Swedish Research Council and the National Space Board in Sweden. Additional support from INAF in Italy and CNES in France for science analysis during the operations phase is also gratefully acknowledged. This research is partially supported by NASA through the Fermi Guest Investigator Grants NNX09AT92G and NNX10AP22G. \\

This work made use of data supplied by the UK Swift Science Data Centre at the University of Leicester.

\begin{minipage}[t]{0.5\textwidth}
$^{1}$ IFAE, Edifici Cn., Campus UAB, E-08193 Bellaterra, Spain\\
$^{2}$Universit\`a di Udine, and INFN Trieste, I-33100 Udine, Italy\\
$^{3}$INAF National Institute for Astrophysics, I-00136 Rome, Italy\\
$^{4}$Universit\`a  di Siena, and INFN Pisa, I-53100 Siena, Italy\\
$^{5}$Croatian MAGIC Consortium, Rudjer Boskovic Institute, University of Rijeka and University of Split, HR-10000 Zagreb, Croatia\\
$^{6}$Max-Planck-Institut f\"ur Physik, D-80805 M\"unchen, Germany\\
$^{7}$Universidad Complutense, E-28040 Madrid, Spain\\
$^{8}$Inst. de Astrof\'isica de Canarias, E-38200 La Laguna, Tenerife, Spain\\
$^{9}$University of \L\'od\'z, PL-90236 Lodz, Poland\\
$^{10}$Deutsches Elektronen-Synchrotron (DESY), D-15738 Zeuthen, Germany\\
$^{11}$ETH Zurich, CH-8093 Zurich, Switzerland\\
$^{12}$Universit\"at W\"urzburg, D-97074 W\"urzburg, Germany\\
$^{13}$Centro de Investigaciones Energ\'eticas, Medioambientales y Tecnol\'ogicas, E-28040 Madrid, Spain\\
$^{14}$Universit\`a di Padova and INFN, I-35131 Padova, Italy\\
$^{15}$Technische Universit\"at Dortmund, D-44221 Dortmund, Germany\\
$^{16}$Inst. de Astrof\'isica de Andaluc\'ia (CSIC), E-18080 Granada, Spain\\
$^{17}$Universit\`a dell'Insubria, Como, I-22100 Como, Italy\\
$^{18}$Unitat de F\'isica de les Radiacions, Departament de F\'isica, and CERES-IEEC, Universitat Aut\`onoma de Barcelona, E-08193 Bellaterra, Spain\\
$^{19}$Institut de Ci\`encies de l'Espai (IEEC-CSIC), E-08193 Bellaterra, Spain\\
$^{20}$Japanese MAGIC Consortium, Division of Physics and Astronomy, Kyoto University, Japan\\
$^{21}$Finnish MAGIC Consortium, Tuorla Observatory, University of Turku and Department of Physics, University of Oulu, Finland\\
$^{22}$Inst. for Nucl. Research and Nucl. Energy, BG-1784 Sofia, Bulgaria\\
$^{23}$Universitat de Barcelona (ICC, IEEC-UB), E-08028 Barcelona, Spain\\
$^{24}$Universit\`a di Pisa, and INFN Pisa, I-56126 Pisa, Italy\\
$^{25}$now at Ecole polytechnique f\'ed\'erale de Lausanne (EPFL), Lausanne, Switzerland\\
$^{26}$now at Department of Physics \& Astronomy, UC Riverside, CA 92521, USA\\
$^{27}$now at Finnish Centre for Astronomy with ESO (FINCA), Turku, Finland\\
$^{28}$also at INAF-Trieste\\
$^{29}$also at Instituto de Fisica Teorica, UAM/CSIC, E-28049 Madrid, Spain\\
$^{30}$now at: Stockholm University, Oskar Klein Centre for Cosmoparticle Physics, SE-106 91 Stockholm, Sweden\\
$^{31}$now at GRAPPA Institute, University of Amsterdam, 1098XH Amsterdam, Netherlands\\
$^{32}$ Santa Cruz Institute for Particle Physics, University of California, Santa Cruz, CA 95064, USA\\
$^{33}$ W. W. Hansen Experimental Physics Laboratory, Kavli Institute for Particle Astrophysics and Cosmology, Department of Physics and SLAC National Accelerator Laboratory, Stanford University, Stanford, CA 94305, USA\\
$^{34}$ Solar-Terrestrial Environment Laboratory, Nagoya University, Nagoya 464-8601, Japan\\
$^{35}$ Dipartimento di Fisica, Universita' di Trieste and INFN Trieste, I-34127 Trieste, Italy
\end{minipage}

\label{lastpage}

\end{document}